\documentclass[aps,prl,preprint,groupedaddress]{revtex4-1}

\usepackage{hyperref}
\usepackage{graphicx}
\usepackage{dcolumn}
\usepackage{bm}

\begin{document}


\title{Universality Class of the Mott Transition}


\author{M. Abdel-Jawad}
\email[]{jawad@riken.jp}
\affiliation{Condensed Molecular Materials Laboratory, RIKEN (The Institute of Physical and Chemical Research) 2-1, Hirosawa, Wako, Saitama, 351-0198, Japan}
\author{I. Watanabe}

\affiliation{RIKEN-RAL, Nishina Centre, 2-1, Hirosawa, Wako, Saitama, 351-0198, Japan.}

\author{N. Tajima}

\affiliation{Department of Physics, Toho University, Miyama 2-2-1, Funabashi-shi, Chiba 274-8510, Japan.}

\author{Y. Ishii}

\affiliation{Department of Physics, College of Engineering, Shibaura Institute of Technology, 307 Fukasaku, Saitama, 337-8570, Japan}

\author{R. Kato}

\affiliation{Condensed Molecular Materials Laboratory, RIKEN (The Institute of Physical and Chemical Research) 2-1, Hirosawa, Wako, Saitama, 351-0198, Japan}


\date{\today}

\begin{abstract}
Pressure dependence of the conductivity and thermoelectric power is measured through the Mott transition in the layer organic conductor $\mathrm{EtMe_3P[Pd(dmit)_2]_2}$. The critical behavior of the thermoelectric effect provides a clear and objective determination of the Mott-Hubbard transition during the isothermal pressure sweep. Above the critical end point, the metal-insulator crossing, determined by the thermoelectric effect minimum value, is not found to coincide with the maximum of the derivative of the conductivity as a function of pressure. We show that the critical exponents of the Mott-Hubbard transition fall within the Ising universality class regardless of the dimensionality of the system.
\end{abstract}


\maketitle


Mott transitions can be achieved in two ways. Filling-controlled Mott transitions are achieved by adding charge carriers through chemical doping or electric field\cite{Newns1998} while bandwidth-controlled Mott transitions\cite{Imada1998} require the control of the ratio $U/W$ between the on-site Coulomb repulsion, $U$ and the bandwidth, $W$. Bandwidth-controlled Mott transitions can be achieved by chemical pressure or, as in this study, by physical pressure in various media. By performing pressure sweep measurements of the conductivity, $\sigma$ at constant temperature (isothermal) in $\mathrm{(V_{1-\mathit{x}}Cr_\mathit{x})_2O_3}$, Limelette \textit{et al.}\cite{Limelette2003} found that $\sigma$ on the metallic side of a three-dimensional Mott insulator could be scaled. They assumed that $\sigma(T,P)-\sigma_c$, where $\sigma_c$ is the critical conductivity at the critical end point, is directly proportional to the order parameter. This implies that $\sigma(T_c,P)-\sigma_c\propto(P-P_c)^{1/\delta}$ and $\sigma(T,P_I(T))-\sigma_c\propto(T_c-T)^\beta$ where $T_c$ and $P_c$ are the temperature and pressure at the the critical end point, $P_I$ is the pressure limit of the spinodal insulating phase at temperatures below $T_c$ and $\delta$ and $\beta$ are critical exponents associated with the singular dependence of the magnetization in proximity to the critical end point.  Values\cite{Limelette2003} of $\delta$ = 3 and $\beta$ = 1/2 were found to be consistent with mean-field theory\cite{Georges1996,Kotliar2000}. The same measurement was subsequently performed in $\mathrm{\kappa-(BEDT-TTF)_2Cu[N(CN)_2]Cl}$ \cite{Kagawa2005} (denoted by $\mathrm{\kappa-Cl}$ hereafter) a quasi-two-dimensional (quasi-2D) organic conductor, where BEDT-TTF stands for bis(ethylenedithio)tetrathiafulvalene. This subsequent study showed that although the conductivity on the metallic side of the phase diagram could still be scaled around the critical endpoint; the extracted critical exponents were inconsistent with any of the known universality class. NMR data in $\mathrm{\kappa-Cl}$ \cite{Kagawa2009} further confirmed those anomalous critical exponents while thermal expansion\cite{Bartosch2010} in the same class of compounds found critical exponents that do correspond to 2D Ising universality. Theoretical interpretations of those results are either in support of the unconventional critical exponents values\cite{Imada2005, Imada2010, Sentef2011} or against\cite{Papanikolaou2008,Semon2012}.

Like $\mathrm{\kappa-Cl}$, $\mathrm{EtMe_3P[Pd(dmit)_2]_2}$ (P2$_1$/m phase) is a quasi-2D organic conductor where $\mathrm{Et=C_2H_5}$, $\mathrm{Me=CH_3}$, dmit=1,3-dithiole-2-thione-4,5-dithiolate. With increasing pressure, $\mathrm{EtMe_3P[Pd(dmit)_2]_2}$ undergoes a Mott transition towards a metallic phase\cite{Shimizu2007} which falls within the compressibility region of helium. In this Letter we report, for the first time, the critical nature of the Mott transition in $\mathrm{EtMe_3P[Pd(dmit)_2]_2}$ from resistivity and thermoelectric power data. We show that the Mott transition belongs to the Ising universality class.

High-quality single crystals of $\mathrm{EtMe_3P[Pd(dmit)_2]_2}$ were prepared by the air oxidation of $\mathrm{(EtMe_3P)_2[Pd(dmit)_2]}$ in an acetone solution containing acetic acid at 5 - 10 $^\circ$C for 3 - 4 weeks. Resistivity and thermoelectric power measurements were made successively along the in-plane direction by dc methods. Special care was taken to measure the Seebeck coefficient by measuring at least three temperature gradients, $\Delta T$, to check the linearity of the voltage drop with $\Delta T$. We have also limited the maximum temperature gradient to roughly 0.1 K to ensure no self-heating effects. In addition to the sample shown in this Letter, three other $\mathrm{EtMe_3P[Pd(dmit)_2]_2}$ samples were measured in less detail and showed similar results. The Fermi surface\cite{Kato2004} of $\mathrm{EtMe_3P[Pd(dmit)_2]_2}$ is not predicted to cut the Brillouin zone unlike $\mathrm{\kappa-Cl}$ where the $S_B$ sign is found\cite{Yu1991} to depend on the crystallographic direction of the measurement. Ambient pressure measurements of $S_B$ from $\mathrm{EtMe_3P[Pd(dmit)_2]_2}$ confirmed that only small differences in amplitudes exist between both in-plane directions. The temperature stability of the isothermal pressure sweep was within 10 mK. Pressure values were estimated from a pressure gauge directly connected to the same helium pressure medium as the sample piston.

\begin{figure}
\includegraphics[scale=0.43]{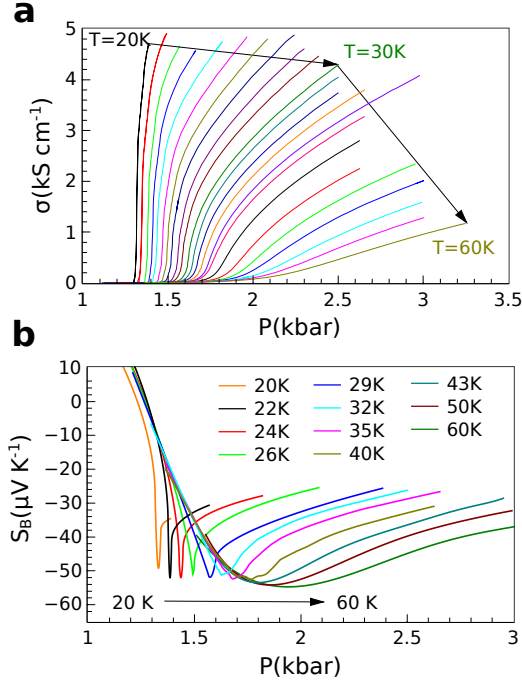}
\caption{\label{fig1} (color online) Limited data set of the isothermal pressure dependence of the conductivity and the Seebeck coefficient from $\mathrm{EtMe_3P[Pd(dmit)_2]_2}$. (a) Pressure dependence of the conductivity. Temperature steps are 1 K between 20 K and 35 K. Above 35 K, the temperatures are 37 K, 40 K, 43 K, 46 K, 50 K, 54 K and 60 K. (b) Pressure dependence of the Seebeck coefficient.}
\end{figure}

 Figure \ref{fig1}(a) shows the isothermal pressure dependence of the conductivity, $\sigma$ in $\mathrm{EtMe_3P[Pd(dmit)_2]_2}$. The first order jump of $\sigma$ is clearly seen at 20 K and is suppressed with increasing temperature until it disappears at the critical temperature, $T_c$ between 29 K and 31 K. We note the absence of clear hysteresis of the first order transition\cite{Supplemental} which we suspect originates from the small temperature distance $\Delta$T $\le$ 1/3 of our measurements from $T_c$. We should note that the high degree of geometrical frustration from this system\cite{Shimizu2007} opens the possibility of a continuous Mott transition as theoretically predicted\cite{Meng2010,Sorella2012}. We have found it difficult to determine objectively the critical pressures of the Mott transition from fittings of $\sigma$ as done in previous studies. To address those difficulties, we have measured the thermoelectric power in the same compound.

\begin{figure*}
\includegraphics[scale=0.45]{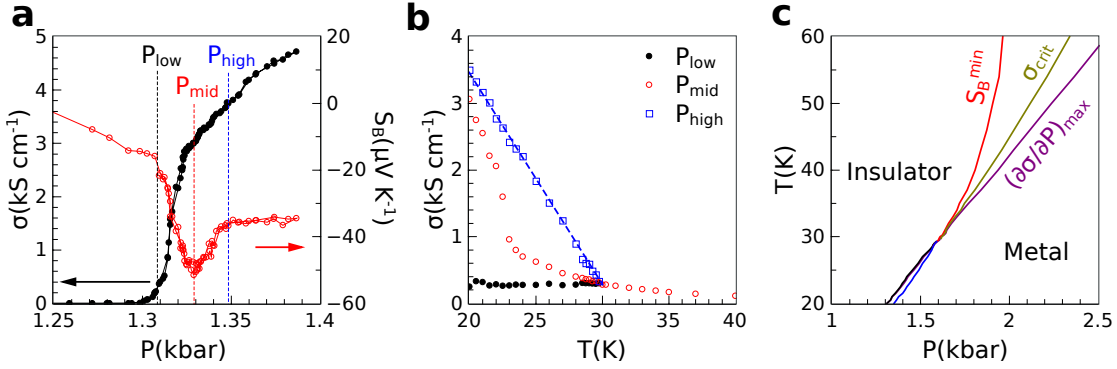}
\caption{\label{fig2} (color online) Phase diagram determination of $\mathrm{EtMe_3P[Pd(dmit)_2]_2}$. (a) Pressure dependence at 20 K of the conductivity and Seebeck coefficient. No appreciable hysteresis was noted. Three significant pressures are highlighted from the curvature change of the Seebeck coefficient pressure dependence and are denoted as $P_\mathrm{low}$, $P_\mathrm{mid}$ and $P_\mathrm{high}$. (b) Temperature dependence of the conductivity along the $P_\mathrm{low}$, $P_\mathrm{mid}$ and $P_\mathrm{high}$ pressures deduced from the Seebeck coefficient pressure dependence. (c) Phase diagram of $\mathrm{EtMe_3P[Pd(dmit)_2]_2}$. Red, dark yellow and purple curves above the critical endpoint mark temperature pressure dependence of the minimum Seebeck coefficient value, the $\sigma_\mathrm{crit}$ = 300 S/cm and the maximum of the pressure derivative of the conductivity respectively. Blue and black curves follow the temperature dependence of $P_\mathrm{low}$ and $P_\mathrm{high}$ deduced from the Seebeck coefficient pressure dependence below the critical endpoint.}
\end{figure*}

\begin{figure}
\includegraphics[scale=0.4]{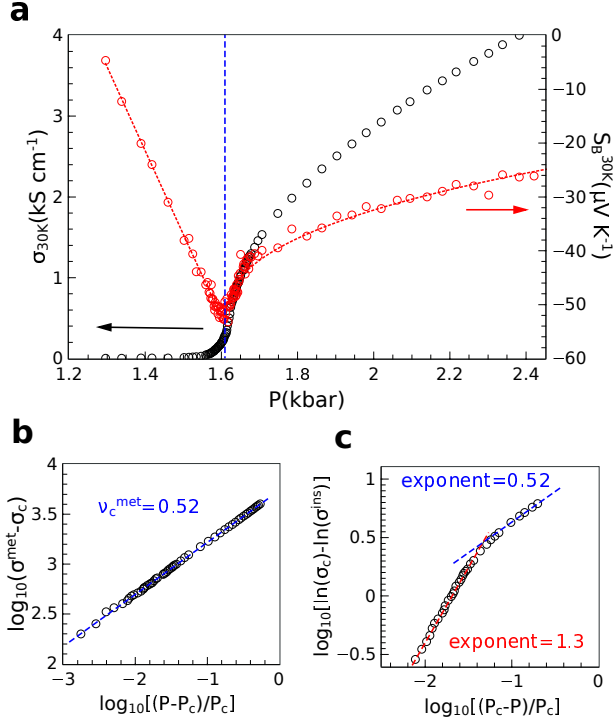}
\caption{\label{fig3} (color online) Scaling at T = 30 K in proximity to the critical temperature $T_c$ $\approx$ 30.15 K of $\mathrm{EtMe_3P[Pd(dmit)_2]_2}$. (a) Pressure dependence of the conductivity and the Seebeck coefficient at $T_c$. Red dotted curve is a least-squares fit of the Seebeck coefficient pressure dependence with the function $S_B^\mathrm{min}+a_\mathrm{met}|P-P_c|^{3\nu_c^\mathrm{met}/4}$ on the metallic side and $S_B^\mathrm{min}+a_\mathrm{ins}|P-P_c|^{2\nu_c^\mathrm{ins}}$ on the insulating side where $S_B^\mathrm{min}$ = -53 $\pm$ 1 $\mu V/K$ is the minimum value of the Seebeck coefficient, $P_c$ = 1612 $\pm$ 20 bar is the critical pressure, $a_\mathrm{met}$ and $a_\mathrm{ins}$ are constants and 3$\nu_c^\mathrm{met}$/4 = 0.39 $\pm$ 0.02 and 2$\nu_c^\mathrm{ins}$ = 1.0 $\pm$ 0.05 are the critical isothermal exponents of the correlation length. (b) Log-log plot of the conductivity from (a) on the metallic side. Blue dotted curve is a straight line with slope of 0.52. (c) Log-log plot of the natural logarithmic function of the conductivity from a on the insulating side. Red and blue dotted curves are straight lines with slopes of 1.3 and 0.52, respectively.}
\end{figure}

	Figure \ref{fig1}(b) shows the isothermal pressure dependence of the thermoelectric power, $S_B$ along the in-plane direction in $\mathrm{EtMe_3P[Pd(dmit)_2]_2}$. $S_B$ exhibits a sharp minimum in its pressure dependence at the Mott transition. The origin of this curvature change is related to the hole character of the gap following the 1/$T$ temperature dependence, as predicted from the Mott formula\cite{Mott2012} in semiconductors\cite{Supplemental}, with a negative metallic background value. Why the metallic value of $S_B$ is negative in this hole system is unclear, but its absolute value follows the standard behavior of a metallic system by decreasing with increasing pressure and conductivity. Preliminary Hall coefficient, $R_H$ measurement\cite{Supplemental} shows strictly positive values which is consistent with dynamic mean-field calculation of strongly correlated metals\cite{PhysRevB.61.7996} where the negative $S_B$ minimum coincided with a coherence crossover. Below the critical end point, the pressure dependence of $S_B$ is found to have three distinct curvature changes which we denote as $P_\mathrm{low}$, $P_\mathrm{mid}$ and $P_\mathrm{high}$ in Fig. \ref{fig2}(a). With increasing temperature, all three pressures join at $P_c$ = 1609 $\pm$ 20 bar and $T_c$ = 30.15 $\pm$ 0.5 K which coincide with the roughly estimated critical end point temperature from the conductivity data. To distinguish which pressures are of importance, we show the temperature dependence of $\sigma$ at those three pressures in Fig. \ref{fig2}(b). We can see that $\sigma(T, P_\mathrm{low})$ is almost temperature independent below $T_c$ with a value equal to the critical conductivity, $\sigma_c$ $\approx$ 300 $\pm$ 20 S/cm at $T_c$. We highlight that $\sigma_c$ value is close to 260 S/cm, the Ioffe-Regel\cite{Ioffe1960} limit for this system\cite{Supplemental}. This limit corresponds to the minimum value of the mean free path of the carriers for coherent transport to occur which cannot be shorter than the unit cell distance. The temperature dependence of $\sigma(T, P_\mathrm{mid})$ does not follow any scalable pattern unlike $\sigma(T, P_\mathrm{high})$ which follows a $T$-linear dependence. From this, we adopt $P_\mathrm{high}$ to estimate the critical pressure, $P_I$ between the spinodal region and the metallic state below $T_c$. Figure \ref{fig3}(a) shows that at $T_c$ a single sharp minimum is the only feature of $S_B$. This minimum broadens as temperature is increased and marks a crossover pressure, $P_\mathrm{cross}^{S_B}$ of the Mott transition above the critical end point. If $\sigma$ is indeed directly proportional to the order parameter of the Mott transition then $P_\mathrm{cross}^{\sigma(T)}$ can be estimated from the maximum of $\sigma'_T(P)\equiv\partial\sigma_T(P)/\partial P$ \cite{Limelette2003}. We compare both estimates of $P_\mathrm{cross}^{S_B(T)}$ and $P_\mathrm{cross}^{\sigma(T)}$ in Fig. \ref{fig2}(c) and find increasing differences with temperature increases above $T_c$. We find that $P_\mathrm{cross}^{S_B(T)}$ can also be deduced from the natural logarithmic function of the conductivity, ln$(\sigma)$ as the maximum of ln$'\sigma_T(P)\equiv \partial$ln$\sigma_T(P)/\partial P$ coincided with the minimum of $S_B$\cite{Supplemental}. It is well known that the maxima from different thermophysical quantities can define different crossover lines emanating from the critical end point\cite{Xu2005,Simeoni2010}. Similar to the case of liquid-gas phase transition\cite{Simeoni2010}, the Widom line which is defined by the specific heat maximum at constant pressure, can be used to delimit the insulatorlike and metal-like limits of the phase diagram above the critical end point with a bad metal region\cite{Gunnarsson2003,Hussey2004} in between both states. 

	To identify the universality class of the Mott transition, we must assign the scaling exponents of $\sigma$ and $S_B$ to functions of the order parameter or the coherence length $\xi$ of the bound states. In an insulator, $\xi$ is also called the localization length and is related to the band gap energy, $\Delta E$ by the relation $\Delta E\simeq h^2/(2m\xi^2)$ \cite{0953-8984-15-19-322}, where $m$ is the charge mass and $h$ is Planck’s constant. We begin with the scaling exponents of $S_B$ on the insulating side. As stated previously, $S_B$ is well fitted by the Mott formula in semiconductors. In the absence of polarons, the magnitude of the 1/$T$ term should be equal or close to the Coulomb gap value found from the conductivity data. Ambient pressure data confirms the absence of polarons \cite{Supplemental} enabling us to assign the scaling exponents of $S_B$ to $\Delta E$ and $1/\xi^2$. Figure \ref{fig3}(a) shows that the critical isothermal pressure dependence of $S_B$ follows a $P$-linear dependence on the insulating side. As the critical isothermal exponent of $1/\xi$ is $\nu_c$, this implies that $\nu_c$ $\approx$ 0.5, a value close to the 8/15 value of the two-dimensional Ising universality class\cite{Andrea2002}. The critical isotherm dependence of the Seebeck coefficient on the metallic side, $S_B^\mathrm{met}$ is shown in Fig. \ref{fig3}(a) and follows a $(P-P_c)^{0.39}+S_B^\mathrm{crit}$ function. We do not have a theoretical explanation for this pressure dependence as the origin of the negative sign of the metallic $S_B$ is also unclear. Nevertheless, critical exponents deduced from the scaling of $S_B^\mathrm{met}$, shown in Fig. 4(b), are consistent with 2D Ising universality if $S_B^\mathrm{met}-S_B^\mathrm{crit}$ $\propto$  $1/\xi^{3/4}$ is assumed.

\begin{figure}
\includegraphics[scale=0.38]{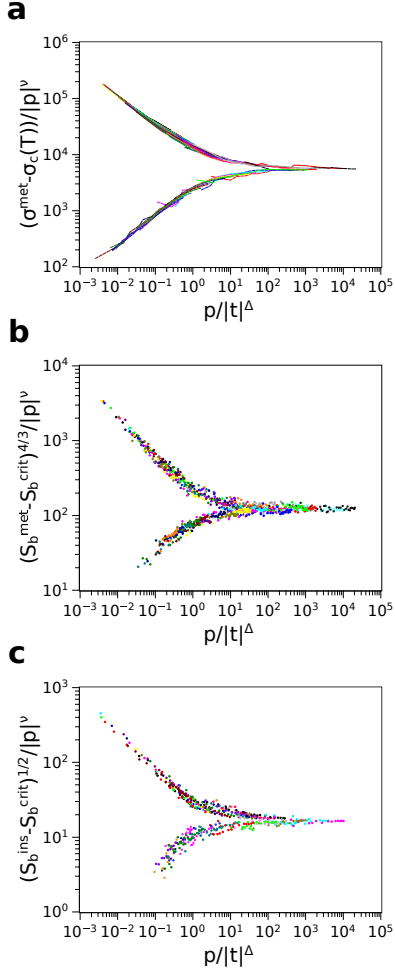}
\caption{\label{fig4} (color online) Scaling plot of the conductivity and Seebeck coefficient of the Mott transition. (a) Plot containing all data on the metallic side of $[\sigma(T, P)-\sigma_c(T)]/|P-P_c|^{\nu_c}$ vs. $[P-P_c(T)]/|T-T_c|^\Delta$. (b) Plot containing all data on the metallic side of $[S_B(T, P)-S_B^\mathrm{crit}]^{4/3}/|P-P_c|^{\nu_c}$ vs. $[P-P_c(T)]/|T-T_c|^\Delta$. Values of $P_c$ in (b) are the exact same as (a). (c) Plot containing all data on the insulating side of $[S_B(T, P)-S_B^\mathrm{crit}]^{1/2}/|P-P_c|^{\nu_c}$ vs. $[P-P_c(T)]/|T-T_c|^\Delta$. Scaled data are the same as in Fig. 1 but with the additional temperatures. Temperatures are 20 K to 24 K by 0.5 K steps, 25 K to 28 K by 1 K steps, 28.5 K to 31 K by 0.25 K steps, 32 K to 35 K by 1 K steps and 37 K, 40 K, 43 K, 46 K, 50 K, 54 K and 60 K. In all scaling plot figures, the data collapse onto two universal curves delimited by temperatures being above or below the critical temperature $T_c$. Values of $T_c$ = 30.15 K, $\nu_c$ = 0.5333 and $\Delta$ = 15/8 are the same in all scaling plot figures. $P_c(T\ge T_c)$ is defined by $\sigma(T,P)$ = $\sigma_c$ = 300 S/cm on the metallic side while on the insulating side it is defined by the minimum of $S_B$ as shown in Fig. 2(c).  $S_B^\mathrm{crit}$ = -52.5 $\mu V/K$ is the same value on either side of the Mott transition.}
\end{figure}	
	
	Scaling of the critical isotherm $\sigma_\mathrm{met}$ is shown in Fig. \ref{fig3}(b) and highlights a square root $P$-dependence. If we assume that $\sigma_\mathrm{met}$ $\propto$ $1/\xi$  then $\nu_c$ is again close to the 8/15 value predicted from the two-dimensional Ising universality class. This relation is consistent with disordered systems scaling\cite{Lee2000} which is not affected by Coulomb correlation\cite{Mott1990}. The role of disorder in a Mott transition is not yet well understood but appears to be important with relation to the conductivity value at the critical end point $\sigma_c$ $\approx$ $\sigma_\mathrm{Ioffe-Regel}$ and the density of state as noted in the $\kappa$ salts \cite{STMLang}. Preliminary Hall coefficient data\cite{Supplemental} shows that in the metallic state, $R_H$ saturates very quickly to a value equivalent to 1 charge per dimer site and highlights that the scaling of $\sigma_\mathrm{met}$ is dominated by the carrier mobility. The scaled critical isothermal dependence on the insulating side of ln$(\sigma_\mathrm{ins})$ is shown in Fig. \ref{fig3}(c) and does not follow a single $P$ dependence, implying that scaling $\sigma_\mathrm{ins}$ is not trivial.  
	
	Scaling theory states\cite{Kadanoff1967} that within proximity to the critical end point, the free energy and the correlation function can be described from a single homogeneous function of the form $\xi(t,p)\simeq p^{-\nu_c} g_\xi^\pm(p/t^\Delta)$ where $t=(T-T_c)/T_c$ is the reduced temperature, $p=(P-P_c)/P_c$ is the reduced pressure, $g_\xi^+$ for $t>0$ and $g_\xi^-$ for $t<0$ are the scaling functions, $\Delta=\sigma\beta$ is the gap exponent and  $\nu=\nu_c\Delta$ is the exponent describing the temperature divergence at $t$ = 0 of $\xi$ along the critical line where $p$ = 0. We have scaled $\sigma_\mathrm{met}$, $S_B^\mathrm{ins}$ and $S_B^\mathrm{met}$ by plotting those functions so that $1/(\xi(t, p)p^{\nu_c})$ is expressed as a function of $p/t^\Delta$. Figure \ref{fig4} shows that such a scaling is possible with values of $\Delta$ = 15/8 and $\nu_c$ = 0.5333 as expected from a 2D Ising universality class. 
	
	Our results show that the Mott transition of $\mathrm{EtMe_3P[Pd(dmit)_2]_2}$ belongs to the 2D Ising universality class. If we assign, as in this study, $\sigma_\mathrm{met}$ to be proportional to $1/\xi$ and not directly to the order parameter then previous results of the Mott criticality yield $\nu_c$ $\approx$ 1/2 and $\nu$ $\approx$ 1 for $\kappa$-Cl \cite{Kagawa2005} and $\nu_c$ $\approx$ 1/3 and $\nu$ $\approx$ 1/2 for $(V_{1-x}Cr_x)_{2}O_3$ \cite{Limelette2003} which are, respectively, the critical exponents of the 2D Ising universality class and the mean-field theory\cite{Andrea2002}. The combined results show that the Mott transition belongs to the Ising universality class regardless of the dimensionality of the system. We should note that it is surprising that within the large range of pressure and temperature of this measurement in $\mathrm{EtMe_3P[Pd(dmit)_2]_2}$  and in $\kappa$-Cl \cite{Kagawa2005}, that no crossover from 2D Ising scaling towards mean-field scaling is found as we move away from the critical end point. 
\begin{acknowledgments}
We thank Professor Andr\'{e}-Marie Tremblay and Professor Ichiro Terasaki for valuable discussions and Professor Yoshiya Uwatoko for his advise on high-pressure experimental techniques. This work was partially supported by Grants-in-Aid for Scientific Research (S) (No. 22224006) from the Japan Society for the Promotion of Science (JSPS).

\end{acknowledgments}

\bibliography{Universality_biblio.bib}

\end{document}